\documentclass{aa}
\usepackage{graphicx}
\usepackage{txfonts}
\usepackage{lscape}

\usepackage{natbib}
\bibpunct{(}{)}{;}{a}{}{,} % to follow the A&A style

\begin{document}
   \title{Transit timing analysis of the exoplanets TrES-1 and TrES-2}

   \author{M. Rabus
          \inst{1,2}
	  \and
	  H. J. Deeg\inst{1,2}
	  \and
          R. Alonso\inst{3,4}
	  \and
	  J. A. Belmonte\inst{1,2}
	  \and
	  J. M. Almenara\inst{1,2}
          }

   \offprints{mrabus@iac.es}

   \institute{Instituto de Astrof\'{i}sica de Canarias,
              V\'ia Lactea s/n, E-38205 La Laguna, Spain\\
        \and
              Departamento de Astrof\'isica, Universidad de La Laguna, 
              E-38205 La Laguna, Tenerife, Spain\\
         \and
             Laboratoire d'Astrophysique de Marseille (UMR 6110),
             Technop\`ole de Marseille-Etoile, F-13388 Marseille, France\\
         \and 
              Observatoire de Gen\`{e}ve, Universit\'{e} de Gen\`{e}ve,
              1290 Sauverny, Switzerland\\
             }

   \date{Received ; accepted }

% \abstract{}{}{}{}{} 
% 5 {} token are mandatory
 
  \abstract
   {}
   {The aim of this work is a detailed analysis of transit light curves from TrES-1 and TrES-2, obtained over a period of three to four years, in order to search for variabilities in observed mid-transit times and to set limits for the presence of additional third bodies.}
   {Using the IAC 80cm telescope, we observed transits of TrES-1 and TrES-2 over several years. Based on these new data and previously published work, we studied the observed light curves and searched for variations in the difference between observed and calculated (based on a fixed ephemeris) transit times. To model possible transit timing variations, we used polynomials of different orders, simulated O-C diagrams corresponding to a perturbing third mass and sinusoidal fits. For each model we calculated the $\chi^2$ residuals and the False Alarm Probability (FAP).}
   {For TrES-1 we can exclude planetary companions ($>$ 1 M$_{\oplus}$) in the 3:2 and 2:1 MMRs having high FAPs based on our transit observations from ground. Additionally, the presence of a light time effect caused by e. g. a 0.09 M$_\odot$ mass star at a distance of 7.8 AU is possible. As for TrES-2, we found a better ephemeris of T$_c = 2\,453\,957.63512(28) + 2.4706101(18) \times$Epoch and a good fit for a sine function with a period of 0.2~days, compatible with a moon around TrES-2 and an amplitude of 57~s, but it was not a uniquely low $\chi^2$ value that would indicate a clear signal. In both cases, TrES-1 and TrES-2, we were able to put upper limits on the presence of additional perturbers masses. We also conclude that any sinusoidal variations that might be indicative of exomoons need to be confirmed with higher statistical significance by further observations, noting that TrES-2 is in the field-of-view of the Kepler Space Telescope.}
   {}


   \maketitle
%
%________________________________________________________________

\section{Introduction}

In 1992, the first exoplanets around the pulsar PSR B1257+12 were discovered by detecting anomalies in the pulsation period \citep{1992Natur.355..145W}. Similarly, by studying variations in the time of transit occurrence, transiting exoplanets give another possibility to find additional companions, even down to Earth masses. There are different mechanisms causing these variations. For one the gravitational influence of a perturbing body can alter the orbital period of the transiting planet \citep{2005Sci...307.1288H,2005MNRAS.359..567A}. This effect is particularly sensitive to detect additional bodies in mean-motion resonances with the transiting planet or to detect moons around that planet. For another, a perturbing mass in an orbit larger than the transiting planet can cause the ``star - transiting planet'' system to wobble around the barycenter and alter the observed periodicity, something that is known as the light-time effect \citep{1959AJ.....64..149I,1995EM&P...71..153S, 2004IAUS..213...80D,2005ASPC..335..191S}. These anomalies are reflected in the times of the transit occurrence. Hence, using a linear ephemeris and several observations of transits, it is possible to get the difference 'O-C' between the observed transit times and the calculated ones from the ephemeris. From this difference we can learn about perturbations on strict orbital periodicity due to a possible companion object which has not been detected yet.  \\
Motivated by the possible detection of low-mass companions around transiting planets, we started a long-term campaign to observe transits of the exoplanets TrES-1 \citep{2004ApJ...613L.153A} and TrES-2 \citep{2006ApJ...651L..61O} over several years. Here we present the results of our observations and a study of the transit timing variations that were found. \\
Searches for planet-mass objects from the light time effect were first proposed for low-mass eclipsing binaries by \citet{1998ASPC..134..224D} and have been performed in-depth on the system CM-Draconis, with ambiguous results to date \citep{2000A&A...358L...5D,2008A&A...480..563D,2009ApJ...691.1400M}. They were soon followed by several studies of timing effects in transiting planet systems: TrES-1 \citep{2005MNRAS.364L..96S}, HD 189\,733 \citep{2008ApJ...682..593M}, HD 209\,458 \citep{2007MNRAS.374..941A,2008ApJ...682..586M}, GJ 436 \citep{2008A&A...487L...5A,2008A&A...487L..25B} and for CoRoT-1b \citep{2009arXiv0903.1845B}. In all cases the authors could only constrain the parameters of a potential second planet.\\
The first search for variability in the transit times of TrES-1 was conducted by \citet{2005MNRAS.364L..96S}. They fitted perturbing planets in different orbits, using transit timing observations from \citet{2005ApJ...626..523C}. \citet{2005MNRAS.364L..96S} did not find any strong evidence of a third object. \citet{2007ApJ...657.1098W} observed three additional transits of TrES-1; they found that the transit times of these three observations occurred progressively later. However, as mentioned by the authors, these three measurements were not enough to make any firm conclusions. Recently, \citet{2009arXiv0905.1833R} reported a transit timing analysis of TrES-1 with the only result of being an improved ephemeris. Previous studies of transit timings for TrES-2 have been performed by \citet{2007ApJ...664.1185H} and \citet{2009arXiv0905.1842R}, neither finding evidence for an unseen perturbing planet. Very recently \citet{2009arXiv0905.4030M} analysed transit time durations between 2006 and 2008, indicating a possible change of inclination.\\
In Section \ref{LC_FIT} we describe the procedure to fit our observations to a transit model and to obtain the observed mid-transit times and their error estimates. These values are then used in Section \ref{TT_an} to interpret the O-C diagram by means of different models. Finally we discuss the results in Section \ref{concl}.\\

%__________________________________________________________________
\section{Light curve fitting}
\label{LC_FIT}

\begin{figure*}
\centering
\includegraphics[angle=90,scale=0.6]{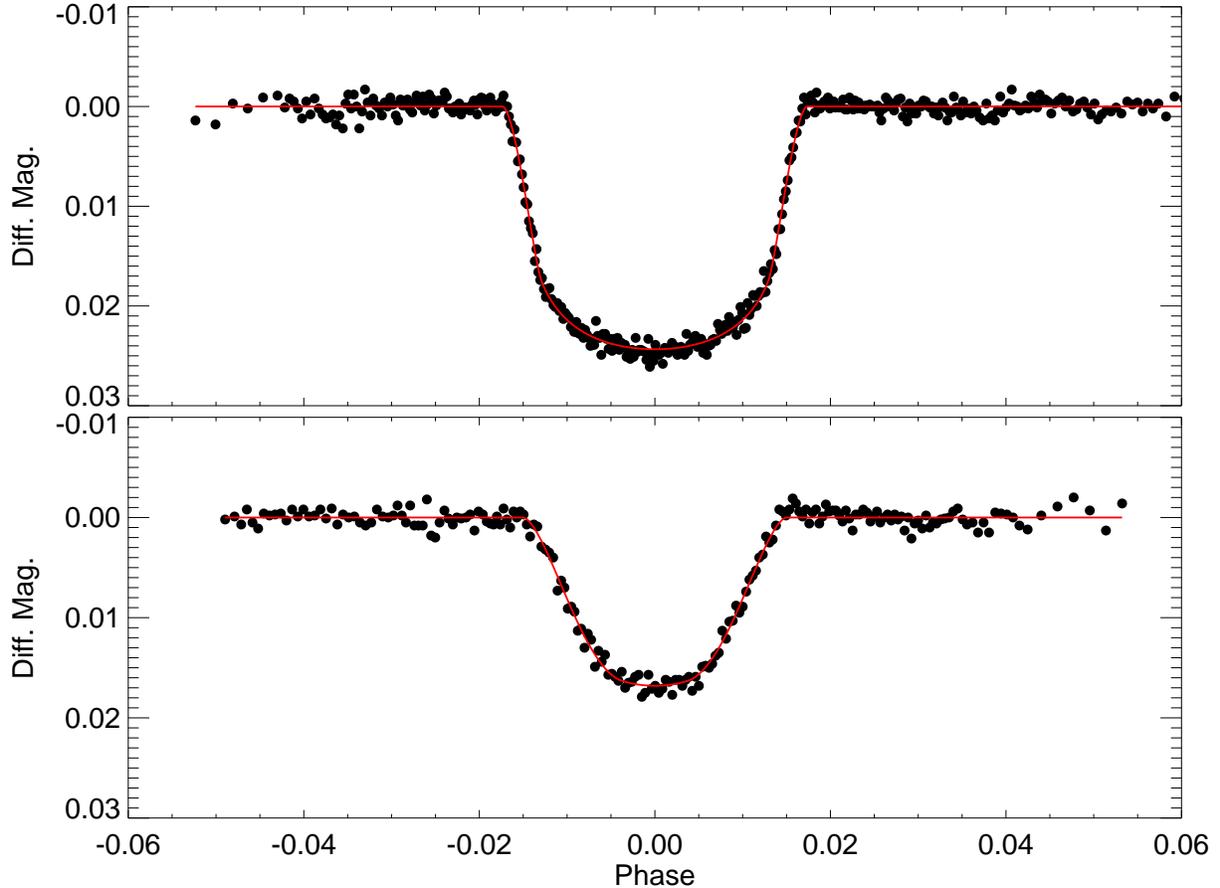}
\caption{Phased and binned light curve of all IAC-80 observations of TrES-1 (top) and TrES-2 (bottom) transits with the solid (red) line being the best-fit model light curve. \label{LC_phased}}
\end{figure*}

\begin{figure}
\centering
\includegraphics[angle=90,scale=0.3]{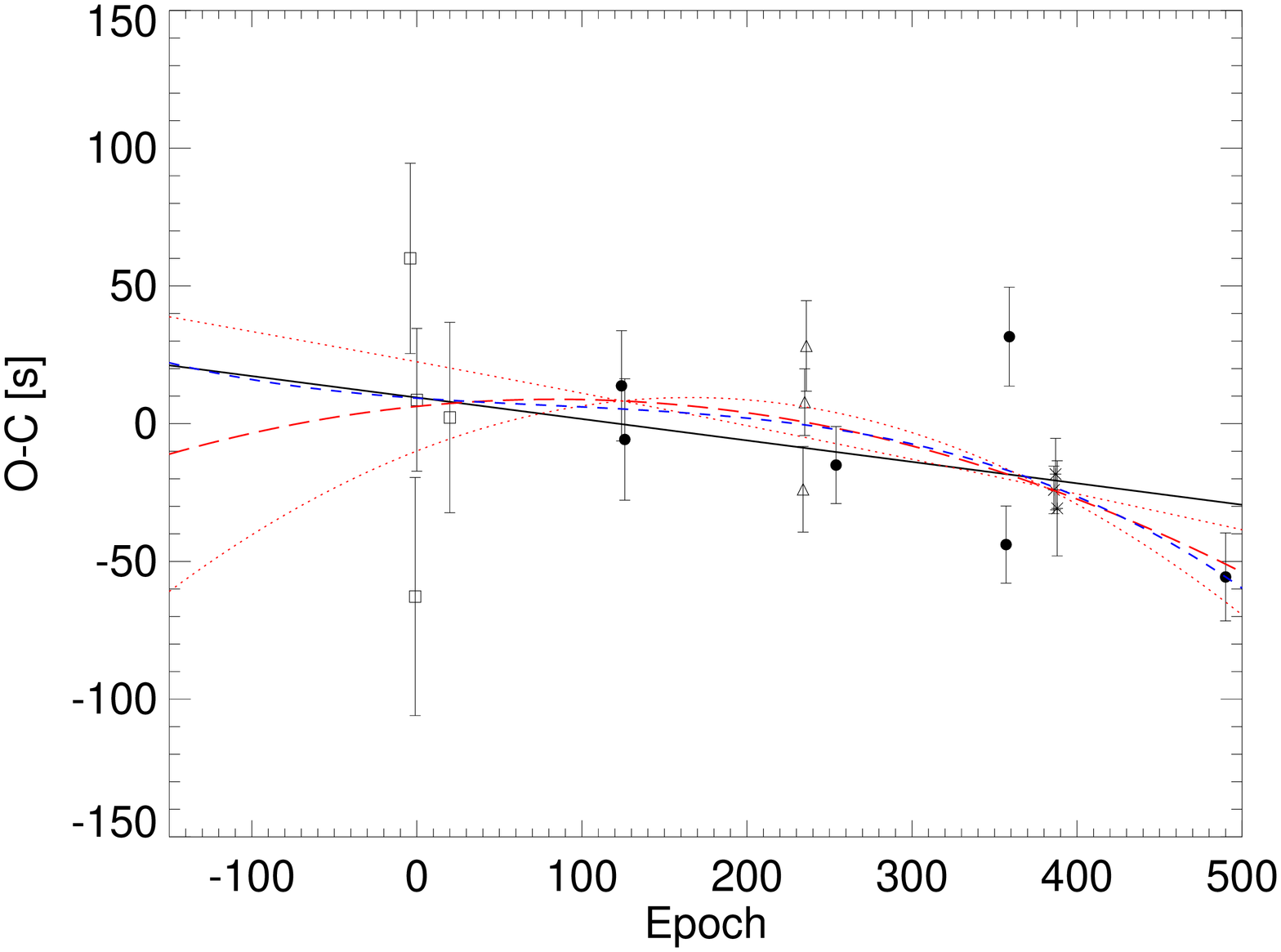}
\caption{Difference between calculated (based on ephemeris from \citet{2007ApJ...657.1098W}) and observed mid-transit times for TrES-1. Filled dots are O-C values obtained with the IAC-80, squares are taken from \citet{2005ApJ...626..523C}, triangles from \citet{2007ApJ...657.1098W} and asterisk are from  \citet{2009IAUS..253..446H}. The lines show different polynomial fits, where the solid black line indicates a linear fit, the long-dashed (red) line a quadratic and the short-dashed (blue) line a cubic polynomial. The dotted (red) lines show the fits corresponding to the variation of the quadratic term within 1-$\sigma$ confidence limits.\label{OC_plot_tres1}}
\end{figure}

\begin{figure}
\centering
\includegraphics[angle=90,scale=0.3]{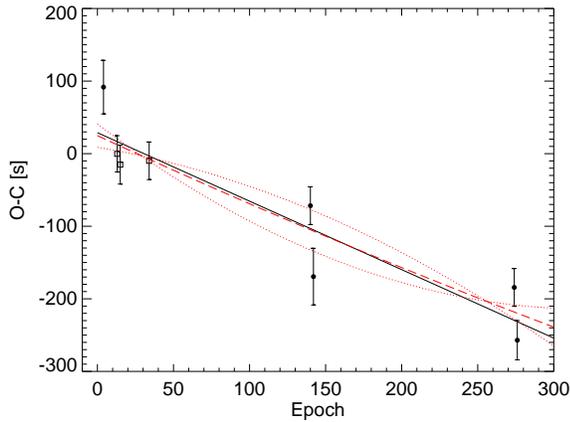}
\caption{Similar as Figure \ref{OC_plot_tres1} (based on ephemeris from \citet{2007ApJ...664.1185H}) including only polynomials fits and corresponding 1-$\sigma$ confidence limits, but for TrES-2. Filled dots are O-C values obtained with the IAC-80 and squares are taken from \citet{2007ApJ...664.1185H}. \label{OC_plot_tres2}}
\end{figure}

All observations reported here were performed with the IAC 80cm telescope (IAC-80) at the Teide Observatory, Tenerife. We used the same observing and analysis procedure for both TrES-1 and TrES-2. For further investigation described here, we only used observations where a complete transit light curve was acquired. A detailed description of the analysis leading from the telescope images to the light curves can be found in \citet{2009A&A...494..391R}, where a study of flux variations during transits due to starspots or additional transiting planets for TrES-1 was presented. We decorrelated the light curves against the airmass by subtracting a quadratic polynomial fit. Due to a slight defocusing of the telescope during observations, the centroids are not well defined and de-correlation against the target's detector position is not applicable. However, the de-focusing and spread of the flux over several pixels will lower the noise correlated with the target's detector position.\\
For TrES-1 we obtained eight useful transit observations (Table \ref{table_tres1}) over a period of three years and for TrES-2 five observations (Table \ref{table_tres2}) over a period of two years. In order to measure the observed mid-transit times, we first created a template of the transit event from a folding in phase and by binning (6-point bins, with a mean size of 66 s for TrES-1 and 90 s for TrES-2) of all observations of the respective transiting planet (Figure \ref{LC_phased}). The standard deviations inside the individual bins were $\sim1.63$~mmag for TrES-1 and $\sim1.79$~mmag for TrES-2, whereas the standard deviation outside the transit part of the binned light curve is 0.7~mmag for TrES-1 and 0.8~mmag for TrES-2. \\
We used these phased and binned light curves to create a model of the transit light curve, using the formalism from \citet{2006A&A...450.1231G} and the simplex-downhill fitting algorithm \citep{1992nrfa.book.....P}. We fitted for the planetary and stellar radii ratio, $k$, the sum of the projected radii, $rr$, and orbital inclination, $i$, while we kept the eccentricity fixed at zero and the limb darkening coefficient fixed, as obtained by the tables of \citet{1995A&AS..114..247C}, assuming quadratic coefficients. The best-fit models are plotted over phased data in Fig. \ref{LC_phased} and the best-fit parameters are shown in Table \ref{param_lc}. We can see that our light curve model parameters are consistent with the parameters from \citet{2008MNRAS.386.1644S} for both TrES-1 and TrES-2.\\ 
We then shifted in time our model with the fixed best-fit parameters against each individual observed light curve and calculated the $\chi^2$ residuals for each shift. The fit with the minimum $\chi^2$ value gave us the observed mid-transit time. We estimated the timing error within a 68~\% (1-$\sigma$) confidence interval of the $\chi^2$ values, given by the range where the $\chi^2$ residuals increases by 1 over the minimum value. We also calculated the timing precision of each individual light curve, $\delta_{t}$, by propagating their respective photometric precision, $\delta_L$, using the equation \citep{2004IAUS..213...80D}:
\begin{equation}\label{deltat0}
\delta_{t}=\delta_L \left[ \sum_i^N \left( \frac{L(t_{i-1})-L(t_{i+1})}{2\Delta t} \right)^2\right]^{-\frac{1}{2}},\\
\end{equation}
where $L(t_i)$ is the stellar brightness at $t_i$, $\Delta t$ is the cadence and the sum goes over all brightness values within the eclipse event. Comparing the estimated error, based on the 1-$\sigma$ interval, with the error calculated using Equation \ref{deltat0} we found a good agreement; both error values are indicated in Tables \ref{table_tres1} and \ref{table_tres2}. Generally the estimated 1-$\sigma$ errors are higher than the propagated errors, but in four cases the error based on the 68~\% confidence interval were lower. Our timing errors are possibly underestimated due to correlated noise. However, we do not modify them in order to stay consistent with the timing measurements and associated errors that we took from the literature.\\
\begin{table}[h]
\caption[]{Comparison between best-fit values of this work and parameters from \citet{2008MNRAS.386.1644S} for transit light curve models of TrES-1 (upper values) and TrES-2 (lower values).}
\label{param_lc}
\centering
\begin{tabular}{ccc}
\hline\hline
\noalign{\smallskip}
 Parameter         & This work &  \citet{2008MNRAS.386.1644S}  \\
\hline
TrES-1 \\
$k$           & 0.1350 & 0.1381 $\pm$ 0.0014    \\
$rr$          & 0.1104 & 0.1097 $\pm$ 0.0022    \\
$i$ [$^\circ$] & 88.67 & 88.67 $\pm$ 0.71    \\
\hline

\noalign{\smallskip}

\hline
TrES-2\\
$k$           & 0.1260 & 0.1268 $\pm$ 0.0032   \\
$rr$          & 0.1462 & 0.1460 $\pm$ 0.0042 \\
$i$ [$^\circ$] & 83.70  & 83.71 $\pm$ 0.42   \\
\hline
\end{tabular}
\end{table}
\begin{table}
\caption[]{Overview of TrES-1 transits used in this work. }
\label{table_tres1}
\centering
\begin{tabular}{cccccc}
\hline\hline
\noalign{\smallskip}
 Epoch & observed         & O-C    & 1-$\sigma$  & Calculated	  & Source     \\
       & transit time & [s]   & error      & timing          &	       \\
       & HJD-2\,450\,000    &        & [s]       & precision [s]	  &	       \\
\hline
  -4    & 3174.6864       & 60   & 35 & -  & 1 \\
  -1    & 3183.7752       & -63  & 43 & -  & 1 \\
   0    & 3186.8061       & 9    & 26 & -  & 1 \\
  20    & 3247.4075       & 2    & 35 & -  & 1 \\
  124   & 3562.5352       & 14   & 20 & 13 & 2 \\
  126   & 3568.5952       & -6   & 22 & 25 & 2 \\
  234   & 3895.8430       & -24  & 16 & -  & 3 \\
  235   & 3898.8734       & 8    & 12 & -  & 3 \\
  236   & 3901.9037       & 28   & 16 & -  & 3 \\
  254   & 3956.4445       & -15  & 14 & 20 & 2 \\
  255   & 3959.4760       & 107  & 17 & 23 & 2 \\
  357   & 4268.5418       & -44  & 14 & 17 & 2 \\
  358   & 4271.5716       & -66  & 17 & 16 & 2 \\
  359   & 4274.6028       & 32   & 18 & 17 & 2 \\
  386   & 4356.4142       & -24  &  9 & -  & 4 \\
  387   & 4359.4443       & -18  & 13 & -  & 4 \\
  388   & 4362.4742       & -31  & 17 & -  & 4 \\
  490   & 4671.5412       & -56  & 26 & 29 & 2 \\
\hline
\noalign{\smallskip}
\noalign{\raggedright Sources: 1) \citet{2005ApJ...626..523C}, 2) This work, 3) \citet{2007ApJ...657.1098W}, 4) \citet{2009IAUS..253..446H}}
\end{tabular}
\end{table}
\begin{table}
\caption[]{Same as Table \ref{table_tres1}, for TrES-2. }
\label{table_tres2}
\centering
\begin{tabular}{cccccc}
\hline\hline
\noalign{\smallskip}
 Epoch & observed         & O-C   & 1-$\sigma$ & Calculated	 & Source\\
       & transit time & [s]   & error     & timing           &	 \\
       & HJD-2\,450\,000    &       & [s]      & precision [s]    &	 \\
\hline
    4   & 3967.5180       & 92   & 37  & 22 & 1 \\
   13   & 3989.7529       & 0    & 25  & -  & 2 \\ 
   15   & 3994.6939       & -15  & 27  & -  & 2 \\ 
   34   & 4041.6358       & -10  & 26  & -  & 2 \\ 
  140   & 4303.5209       & -72  & 26  & 20 & 1 \\
  142   & 4308.4613       & -169 & 39  & 22 & 1 \\
  274   & 4634.5828       & -184 & 26  & 24 & 1 \\
  276   & 4639.5232       & -257 & 27  & 21 & 1 \\
\hline
\noalign{\smallskip}
\noalign{\raggedright Sources: 1) this work, 2) \citet{2007ApJ...664.1185H}}
\end{tabular}
\end{table}
Finally we subtracted the observed mid-transit times from the calculated ones, obtaining the O-C values. The calculated mid-transit times for TrES-1 were obtained from the ephemeris T$_c= 2\,453\,186.8060 + 3.0300737 \times$ Epoch \citep{2007ApJ...657.1098W} and for TrES-2 we used T$_c = 2\,453\,957.6348 + 2.470621 \times$ Epoch \citep{2007ApJ...664.1185H}.\\
In the following transit timing analysis we included several published mid-transit observations, by \citet{2005ApJ...626..523C}, \citet{2007ApJ...657.1098W} and \citet{2009IAUS..253..446H} for TrES-1 (Table \ref{table_tres1}), and for TrES-2 by \citet{2007ApJ...664.1185H} (Table \ref{table_tres2}). We considered only O-C times with errors below 60 sec, which led to the rejection of some O-C values from \citet{2005ApJ...626..523C}. In the O-C residuals of TrES-1 we also removed two outliers at Epoch 255 and 358 which have been identified by \citet{2009A&A...494..391R} as transits with possible starspots. Several further transit times of TrES-2 and TrES-1 were recently reported by \citet{2009arXiv0905.1842R,2009arXiv0905.1833R}, respectively, and for TrES-2 by \citet{2009arXiv0905.4030M}, but we did not include these in our study, since they had individual errors and an internal scatter several times larger than the data included in this study. The transit mid-time corresponding to Epoch 0 of TrES-2, reported by \citet{2006ApJ...651L..61O} has also been removed due to its high error of $>$60~s. We obtained for TrES-1 an O-C diagram spanning four years, with 16 points (Figure \ref{OC_plot_tres1}) and for TrES-2 a diagram spanning two years, with 8 points (Figure \ref{OC_plot_tres2}).\\

%__________________________________________________________________
\section{Transit timing analysis}
\label{TT_an}

Two mechanisms have been brought forward that may cause deviations of transit times from strict periodicity. For one the light time effect and for another the influence of a third body on the transiting planet's intrinsic periodicity.\\

\subsection{Search for the light time effect}

We first consider the light time effect, where the barycenter of the binary system, in our case the system ``star - transiting planet'', is offset against the barycenter formed by an additional third body. This will cause the light to travel a longer or shorter distance to the observer and hence the observer will see a different period, where the intrinsic period is unknown. \\
For the analysis of the O-C times, we fitted three polynomials of different orders to the O-C values, namely a linear, $OC_{fit}=\kappa_0+E \kappa_1$, a parabolic, $OC_{fit}=\kappa_0+E \kappa_1+E^2 \kappa_2$, and a cubic polynomial, $OC_{fit}=\kappa_0+E \kappa_1+E^2 \kappa_2 +E^3 \kappa_3$, where $E$ is the Epoch number and $\kappa$ the fitted polynomial coefficients. We also analyzed the case of fixed original ephemeris ($OC_{fit}=0$) and the case of maintaining the original period and fitting only for a constant offset in O-C ($OC_{fit}=\kappa_0$). Table \ref{bestfit_poly} shows the obtained best fit coefficient for TrES-1 and TrES-2 respectively, and Figs. \ref{OC_plot_tres1} and \ref{OC_plot_tres2} show a plot of the respective polynomials. In order to quantify the improvement of the different fits, we calculated the False Alarm Probability (FAP), which is the significance level of the fit quality improvement and indicates the probability of making a type I error. Therefore, we first calculated the F-values according to equation:
\begin{equation}\label{signi}
F=\frac{(\chi^2_1-\chi^2_2)/(\nu_1-\nu_2)}{(\chi^2_2/\nu_2)},\\
\end{equation}
where $\chi^2_1$ are residuals from the lower order fit, $\chi^2_2$ are residuals from the higher order fit, and $\nu_{1,2}$ are the corresponding degrees of freedom (Table \ref{bestfit_poly}). We then used the F-value to calculate the significance using the IDL-routine {\tt MPFTEST} from the Markwardt IDL library, which gives the probability for a value drawn from the F-distribution to equal or exceed the given F-value. For TrES-1, FAPs were calculated against the original ephemeris, against the offset-only case and against the linear polynomial. We see that the original ephemeris is unlikely to be the best solution with the linear and quadratic fits being the most likely descriptions, having low $\chi^2$ residuals and a low FAP. For TrES-2 a linear trend in the O-C residuals (Fig. \ref{OC_plot_tres2}) is apparent. The statistical analysis shows no clear preference between a linear or a quadratic polynomial (Table \ref{bestfit_poly}). The distribution of the observations of TrES-2 into three groups, acting as pivot points, does not support fits of orders higher than the quadratic one. Therefore, we fitted no cubic polynomial to the TrES-2 data.\\
As for the linear coefficients $\kappa_0$ and $\kappa_1$ these are without relevant physical meaning, but indicate a slightly different ephemeris than used, where $\kappa_0$ is an offset of the mid-transit time at epoch 0 and $\kappa_1$ a correction to the period of the ephemeris. However, we can give a physical meaning to the quadratic coefficient. The quadratic term ($\kappa_2$), gives the system's acceleration along the line of sight by using the equation \citep{2008A&A...480..563D}:
\begin{equation}\label{accel}
a_{\parallel}=2 \frac{c\kappa_2}{P^2},\\
\end{equation}
where $c$ is the speed of light and $P$ is the observed period. \\
In order to estimate the error of the quadratic term, we stepped through different values of $\kappa_2$, fitting for $\kappa_0$ and $\kappa_1$. The $\kappa_2$ values which increased the best-fit $\chi^2$ residuals by 2.3 gave the 1-$\sigma$ confidence limits (see Table \ref{bestfit_poly}); dotted lines in Figs. \ref{OC_plot_tres1} and \ref{OC_plot_tres2} show the corresponding fits at these limits. Our best fit parabola model for TrES-1 gave a quadratic term of $\kappa_2=-3.6 \pm 3.4 \times 10^{-4}$~s. Solving Eq. \ref{accel} we obtained an acceleration of $a_{\parallel}=-3.2 \pm 3.0 \times 10^{-6}$~m~s$^{-2}$ and $\frac{a_{\parallel}}{c}=-1.1 \pm 1.0 \times 10^{-14}$~s$^{-1}$. The cubic term $\kappa_3$ would indicate a constant change in acceleration; however the cubic fit is also less likely than the quadratic one and will not be further discussed. \\
For the quadratic solution of TrES-2 we obtained a best-fit value of $\kappa_2$ smaller than its error. This high error is consistent with a low significance of the quadratic solution, therefore we support the linear case, with a 1-$\sigma$ upper limit for accelerations of $a_{\parallel} \le 3.2 \times 10^{-5}$~m~s$^{-2}$. \\ 
Considering the clear linear trend of O-C times in Figs. \ref{OC_plot_tres1} and \ref{OC_plot_tres2}, and using the coefficients from the linear fit, we indicate here an improved ephemeris for TrES-1, given by: 
\begin{equation}\label{new_ephem_tres1}
T_c = 2\,453\,186.80611(16) + 3.0300728(6) \times \textrm{Epoch},\\
\end{equation}
where the values in parenthesis give the uncertainty in the last two digits and for TrES-2:\\
\begin{equation}\label{new_ephem_tres2}
T_c = 2\,453\,957.63512(28) + 2.4706101(18) \times \textrm{Epoch}.\\
\end{equation}

\subsection{Search for a perturbation of the intrinsic period}
\label{intrinsic_period}

\begin{figure}
\centering
\includegraphics[angle=90,scale=0.33]{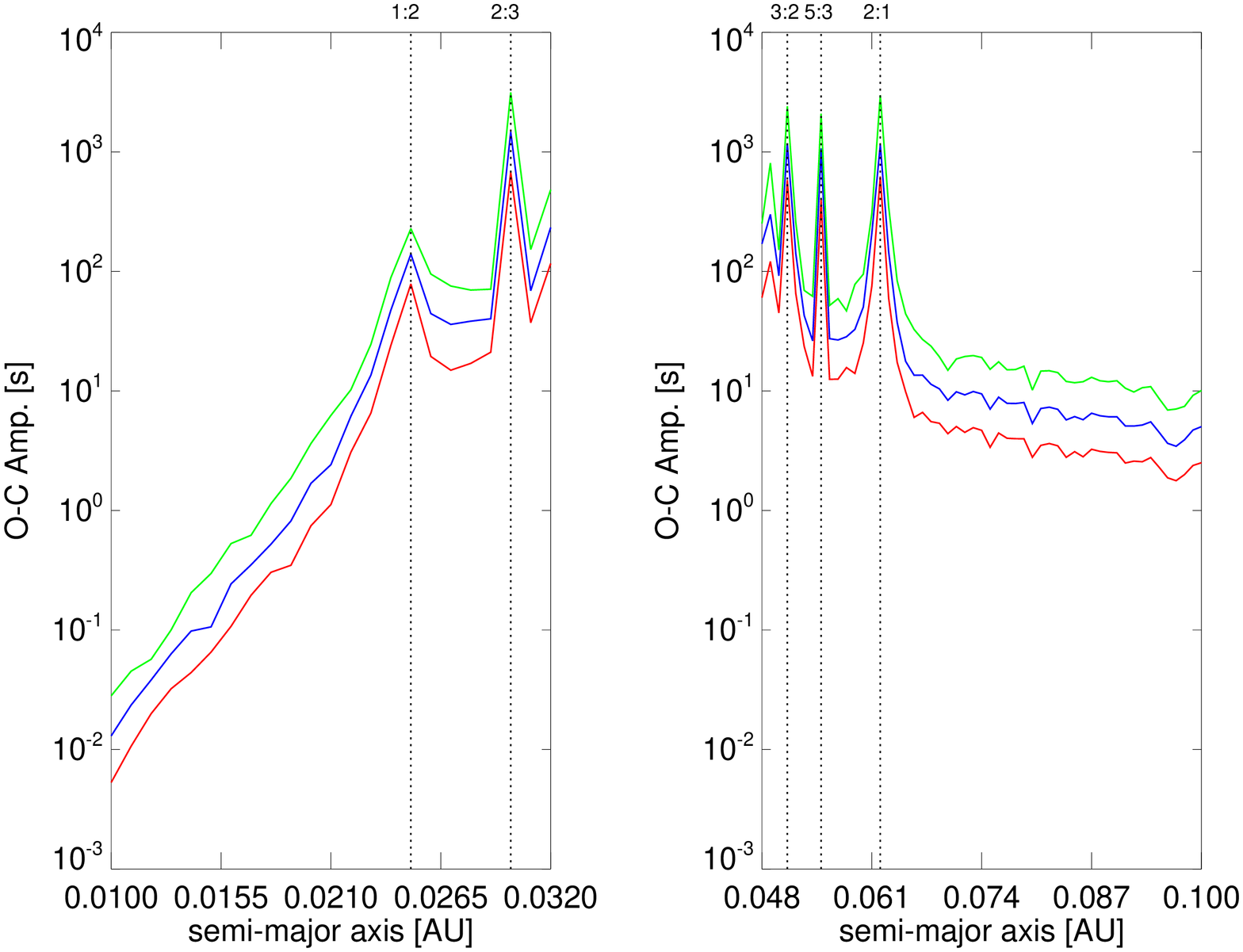}
\includegraphics[angle=90,scale=0.33]{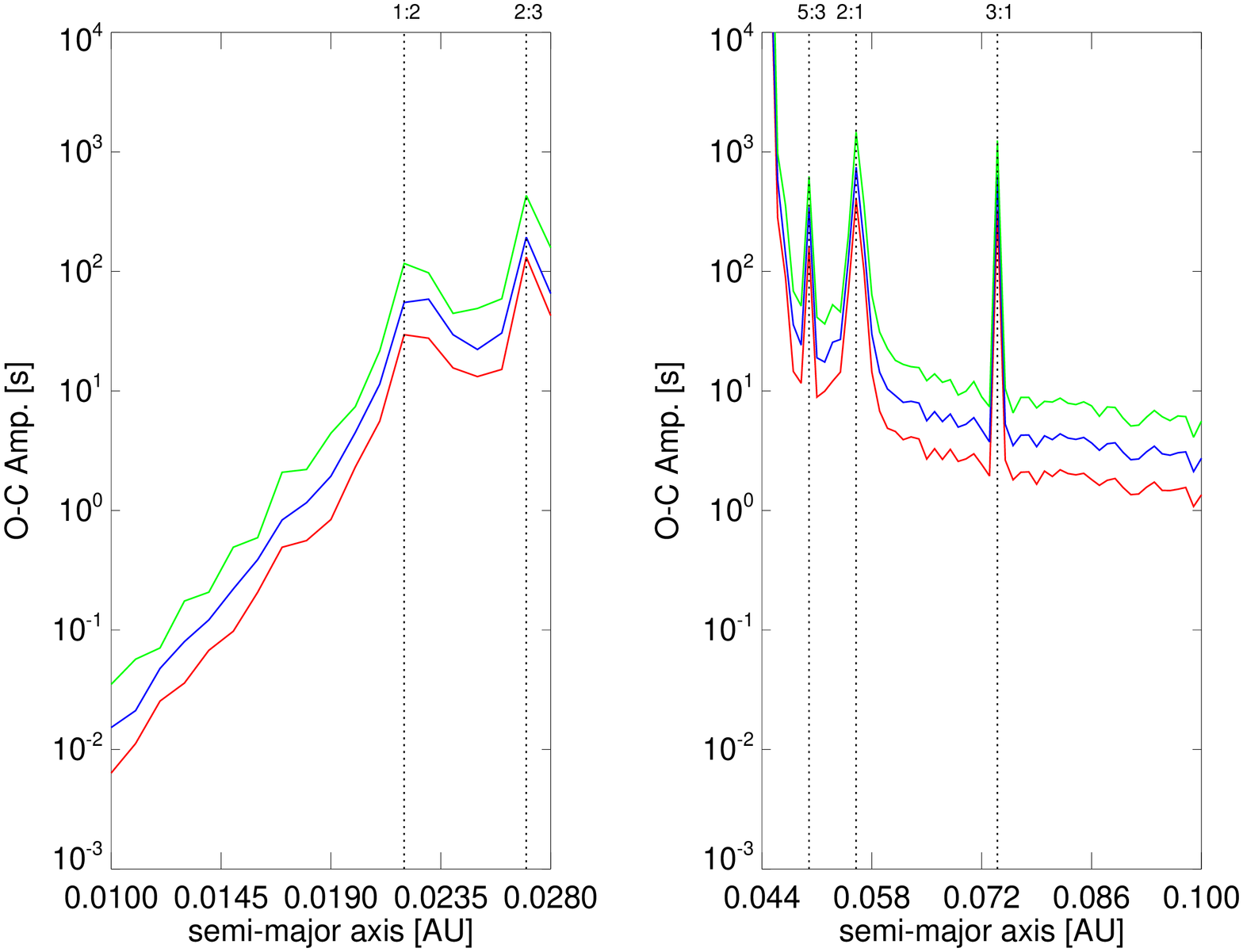}
\caption{Maximum O-C amplitudes of a 2 dimensional N-body simulation for the system TrES-1 (upper two panels) and TrES-2 (lower two panels) due to a perturber during 1000 transits plotted against the semi-major axes of the perturber. The colored solid lines correspond to different perturbing masses, namely 10~M$_\oplus$ (red), 20~M$_\oplus$ (blue) and 40~M$_\oplus$ (green). Left side: inner perturber, right side: outer perturber. Vertical lines indicate the MMRs. \label{OC_sim}}
\end{figure}

\begin{figure}
\centering
\includegraphics[angle=90,scale=0.33]{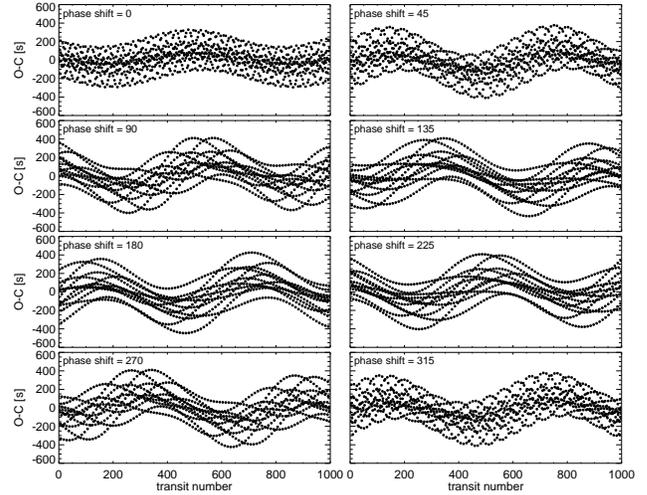}
\caption{Example synthetic O-C diagrams for TrES-1 and a perturber of 30~M$_\oplus$ at 0.053~AU and with different phase shifts between them. Note that the appearance of several lines is due to an aliasing effect between consecutive transits. For a given transit number there is only one O-C time.\label{OC_diag_tres53}}
\end{figure}

\begin{figure}
\centering
\includegraphics[angle=90,scale=0.33]{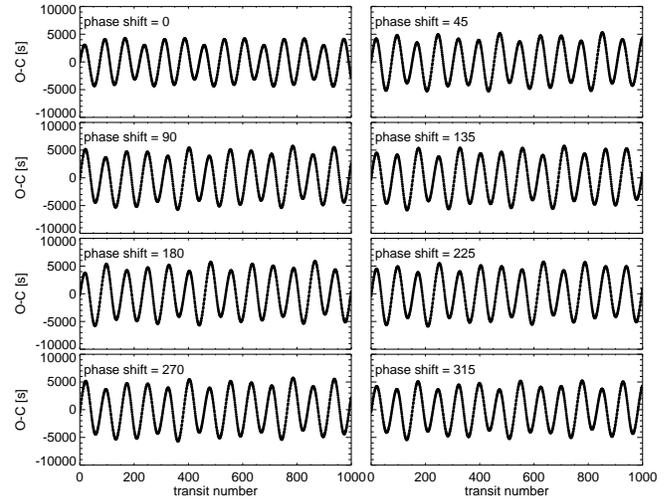}
\caption{Same as Fig. \ref{OC_diag_tres53}, but for a perturber distance of 0.062 AU (2:1 MMR).\label{OC_diag_tres62}}
\end{figure}

\begin{figure}
\centering
\includegraphics[angle=90,scale=0.33]{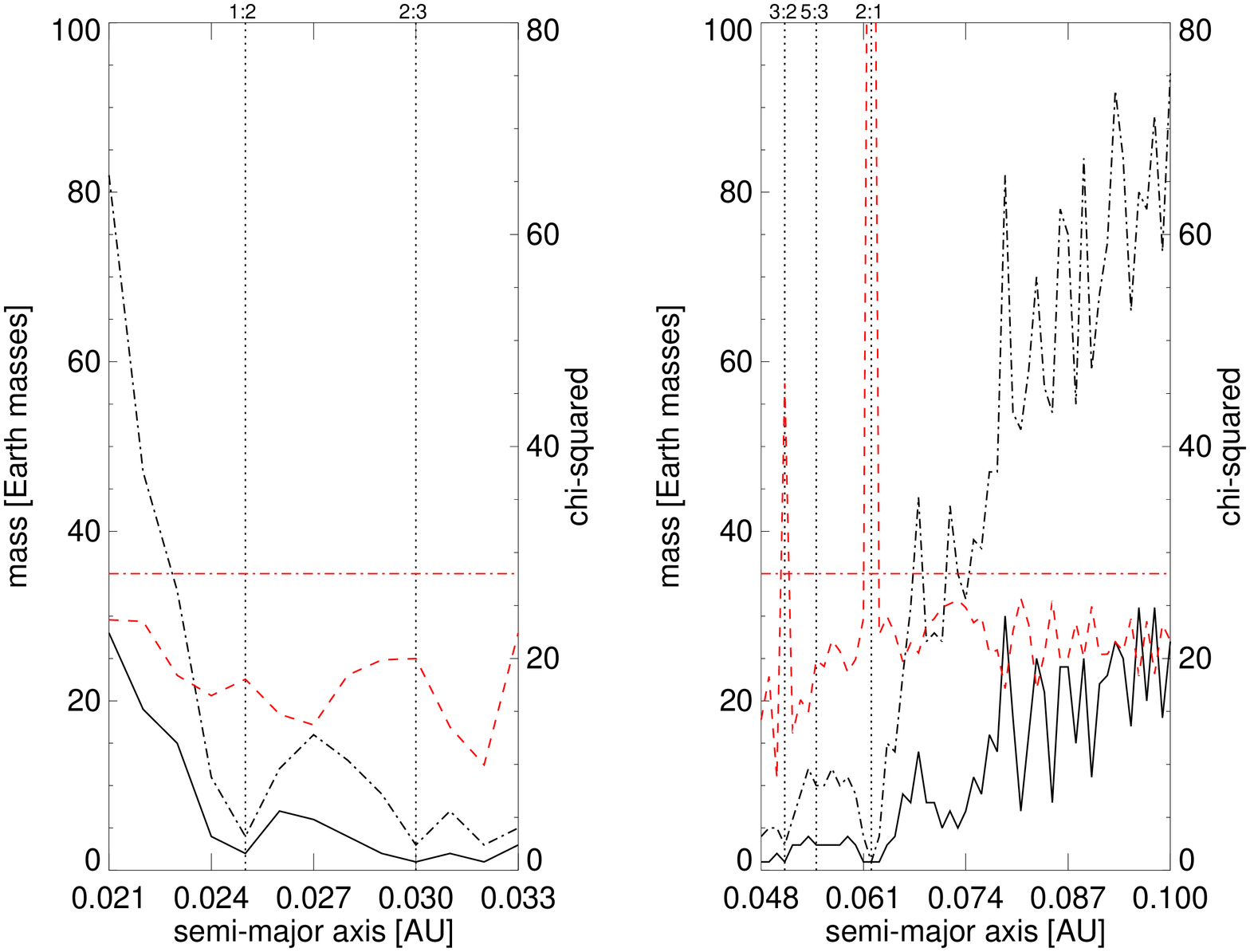}
\includegraphics[angle=90,scale=0.33]{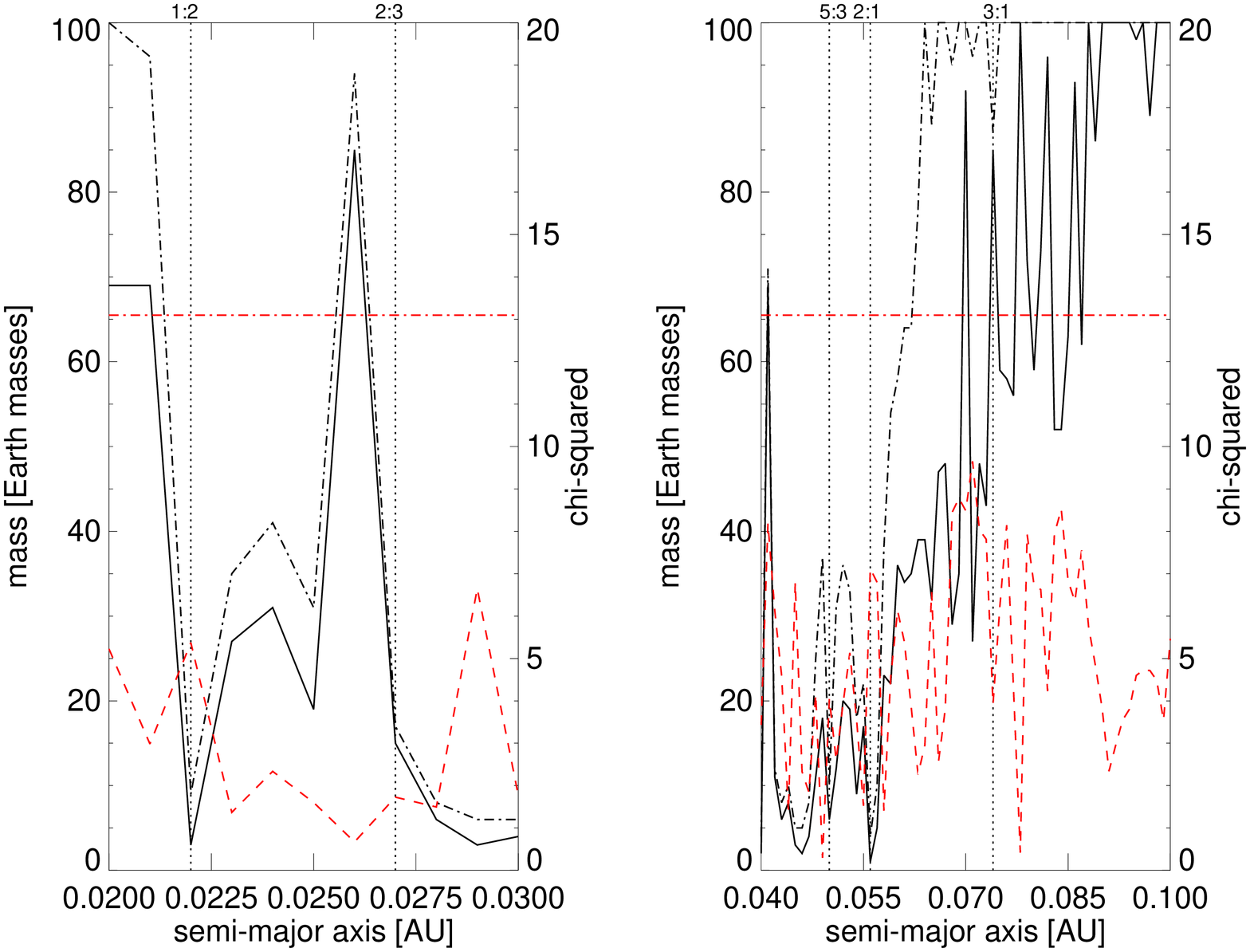}
\caption{Masses of third bodies with the lowest $\chi^2$ (solid black line), based on the best fit of the observed O-C values to the synthetic O-C values for each semi-major axis, and upper mass limits resulting in a FAP of 90\% (dashed-dotted black line) with respect to the linear fit for each semi-major axis. Red lines show the lowest $\chi^2$ values (dashed red line) corresponding to the best-fit masses and the $\chi^2$ value of the linear fit is indicated by a horizontal dashed-dotted red line. The upper plot is for TrES-1 and the lower plot for TrES-2. Mean motion resonances are indicated by vertical dotted lines.\label{bestmass}}
\end{figure}

\begin{figure*}
\centering
\includegraphics[angle=90,scale=0.6]{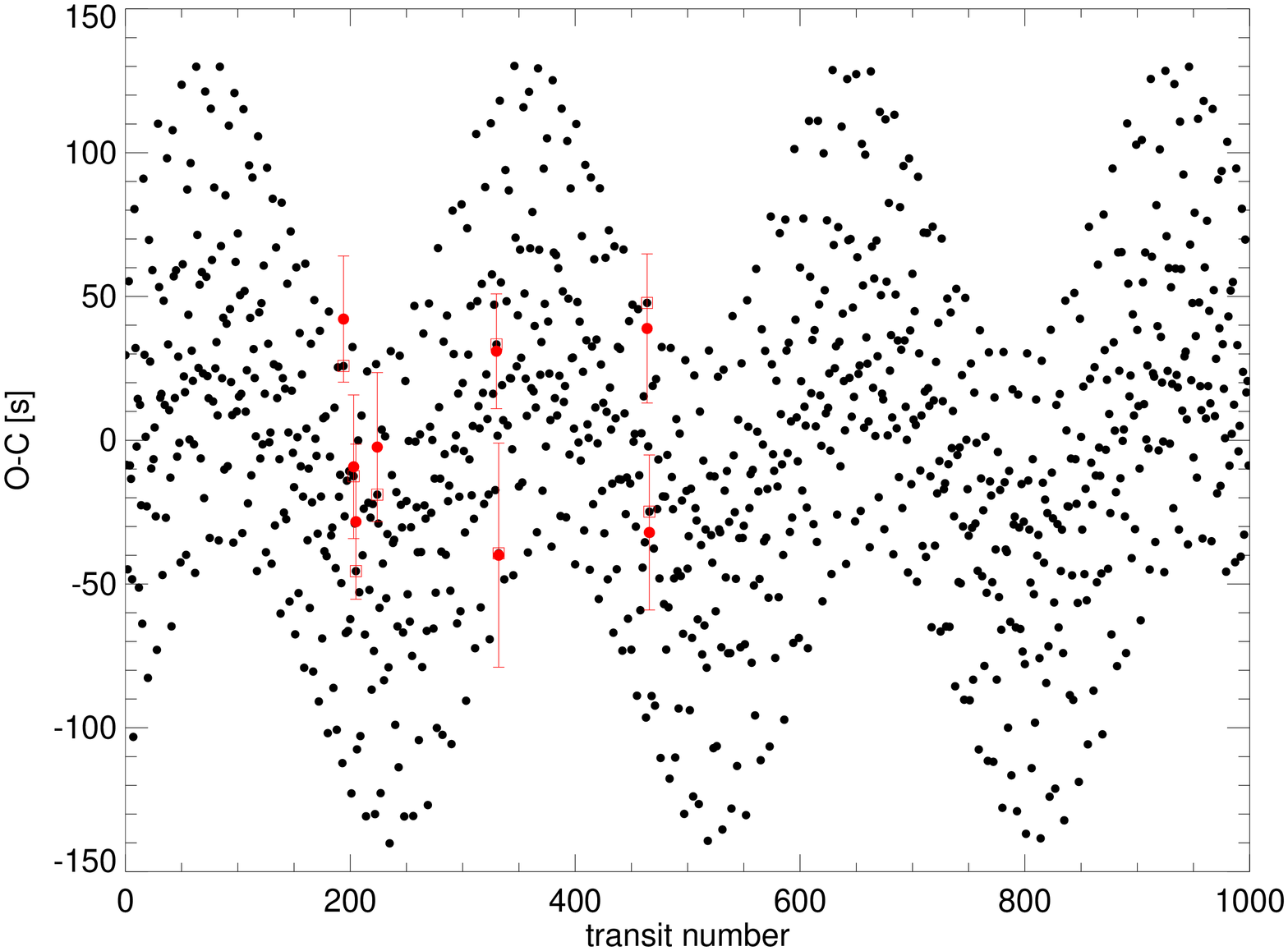}
\caption{Simulated O-C diagram of TrES-2 and the best fit perturber with a mass of 18~M$_\oplus$ and an orbit of 0.051~AU. The red dots correspond to the observations and the squares to the corresponding simulated O-C values. \label{bestfit}}
\end{figure*}

The other cause for transit timing variations could be a perturbation of the intrinsic planet period $P'$ due to the gravitational influence of a third body on the transiting planet. Regarding such perturbations, there exist no analytical equations that describe the gravitational influence on a transiting planet due to an undetected third body. Generally, N-body simulations are used, iterating over a large orbital parameter space for a maximum possible mass range, see e. g. \citet{2005Sci...307.1288H,2005MNRAS.364L..96S,2005MNRAS.359..567A,2007MNRAS.374..941A,2008ApJ...682..586M,2008ApJ...682..593M,2008ApJ...688..636N}.\\
In order to find the parameter space of third bodies compatible with the observed transit times, we created a numerical 2-dimensional simulation of a three body system by integrating over the equations of motion. We considered the problem in two dimensions, assuming that the orbits of the exoplanetary system and its respective perturber are co-aligned. We further considered the problem in a helio-centric frame, with the star in the center of the coordinate system, meaning that we neglect the light-time effect in this context. To integrate the equation of motion we used the Burlisch-Stoer algorithm \citep{1992nrfa.book.....P} with a 1 second time step and an accuracy of 10$^{-10}$. \\
We considered the problem for an inner and an outer perturber separately and neglected the 1:1 mean motion resonance (MMR). Each simulation started with a perturber 0.005 AU away from the transiting planet, with an initial zero eccentricity and with a phase shift towards the transiting planets between 0$^\circ$ and 315$^\circ$ in steps of 45$^\circ$. We advanced the simulations in orbital steps of 0.001~AU and used 100 mass steps from 1~M$_\oplus$ to 100~M$_\oplus$ for the perturbing object. We simulated 1000 transits, corresponding to approximately eight years coverage for TrES-1 and seven years coverage for TrES-2. We used a mass of 0.61~M$_J$ for TrES-1 and a mass of 0.88~M$_{\odot}$ for its host star \citep{2008ApJ...677.1324T}, and a mass of 1.2~M$_J$ for TrES-2 and 0.98~M$_{\odot}$ for its host star \citep{2007ApJ...664.1190S}. After obtaining the simulated O-C diagrams, we applied a Fourier transformation to the synthetic O-C diagrams \citep{2008ApJ...682..586M,2008ApJ...682..593M} and derived the maximum obtained O-C amplitudes as a function of the perturber's semi-major axis; showing them for masses of 10, 20 and 40~M$_{\oplus}$ in Fig. \ref{OC_sim}, where also the MMRs are indicated by vertical lines. Figs. \ref{OC_diag_tres53} and \ref{OC_diag_tres62} show some examples of synthetic O-C diagrams for TrES-1.  \\
To establish potential third body orbital parameters, we fitted for each perturber distance the observed O-C diagram to all synthetic ones that had been generated for masses of 1 - 100~M$_\oplus$ (Fig. \ref{bestmass}). We left three parameters open in the fits: one parameter to shift the O-C diagrams in Epoch, corresponding to find the best moment for the first real observation (E=0) within the simulated data, and two parameters describing linear deviations between simulated and modelled O-C diagrams, which have no relevant physical meaning. \\ 
For the best-fit masses, we see a general trend to higher masses for perturbers closer to the host star and at larger semi-major axis, and we can also identify the mean-motion resonances, where the best fit for a consistent perturber indicates lower masses (vertical lines in Fig. \ref{bestmass}). At larger orbits for the perturber, the best-fit perturber's mass increases and it increases more steeply for TrES-2 than for TrES-1. This is due to TrES-2 having a higher mass than TrES-1 and being closer to the central star. The high $\chi^2$ peaks at the 2:3 and 2:1 MMRs for TrES-1 reveal that a $>$1 M$_{\oplus}$ planet in that configuration might have been detected in our data set.\\ 
Again, in order to quantify the improvement of the simulated O-C fits in the lower MMRs and the best-fit simulated O-C against the linear polynomial, we calculate the FAPs. In these cases it is better not to use Eq. \ref{signi} to calculate the F-value, since this equation is better suited for cases where one expects small changes for additional parameters, whereas for the simulated O-C values, we expect big changes. Therefore, we used $F=\frac{(\chi^2_1/\nu_1)}{(\chi^2_2/\nu_2)}$ to calculate the F-value for the {\tt MPFTEST}-routine. In Table \ref{bestfit_poly}, we show the $\chi^2$ and FAPs for some special cases of our simulated O-C values.\\
For TrES-1 we found that most of the $\chi^2$ values for the simulated O-C models were below the one of the linear fit, but based on the available data, we can exclude the case of 2:3 and 2:1 MMR for planets more massive than 1 M$_\oplus$ with high FAPs. We found a best-fit for a perturber with a mass of 2~M$_\oplus$ and a semi-major axis of 0.05~AU, but this peak is not outstandingly low. Similarly, for TrES-2, all $\chi^2$ values for the O-C models are below the one from the linear fit. We can identify some low $\chi^2$ peaks, with the lowest at 0.049~AU ($\chi^2=0.3$) for a perturbing mass of 18 M$_\oplus$. Fig. \ref{bestfit} shows for TrES-2 the O-C simulation for the lowest $\chi^2$ with the observed O-C residuals over-plotted. Given that none of the simulated perturbers indicate a uniquely low $\chi^2$, we do not find support for any of them. \\
In order to calculate an upper mass limit above which we can reject perturber masses due to a high increase in $\chi^2$, we first set a $\chi^2$ threshold to  50.2 for TrES-1 and 28.5 for TrES-2. These thresholds correspond to a 90 \% FAP against the linear fit using the {\tt MPFTEST}-routine. We then increased the masses for each semi-major axis until the $\chi^2$ residuals reached the previously established threshold, which gave us the upper mass limit. Masses above the threshold can be rejected with a FAP higher than 90 \% (Fig. \ref{bestmass}).

\subsection{Search for a sinusoidal transit timing variability}

\begin{figure}
\centering
\includegraphics[angle=90,scale=0.33]{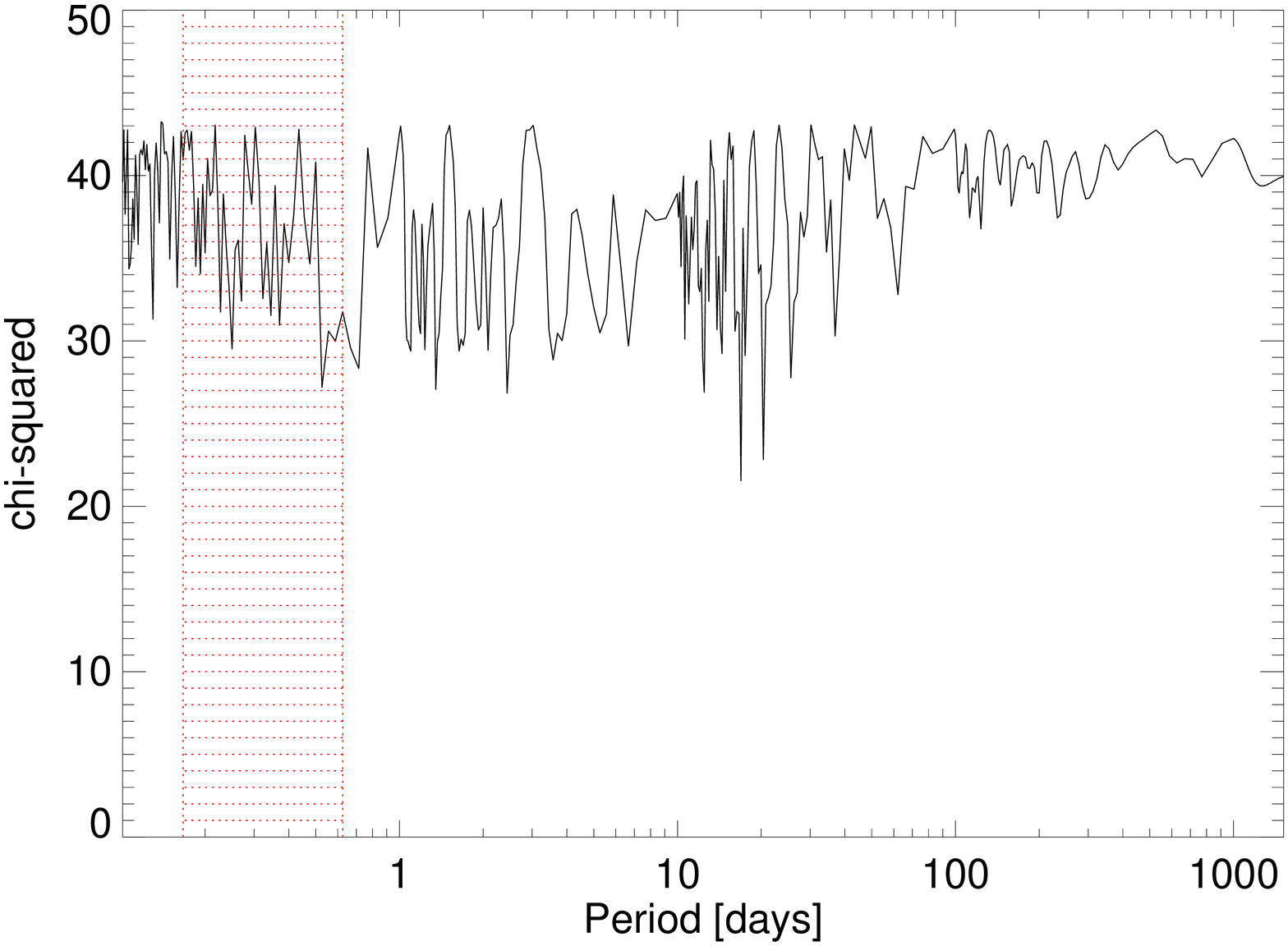}
\includegraphics[angle=90,scale=0.33]{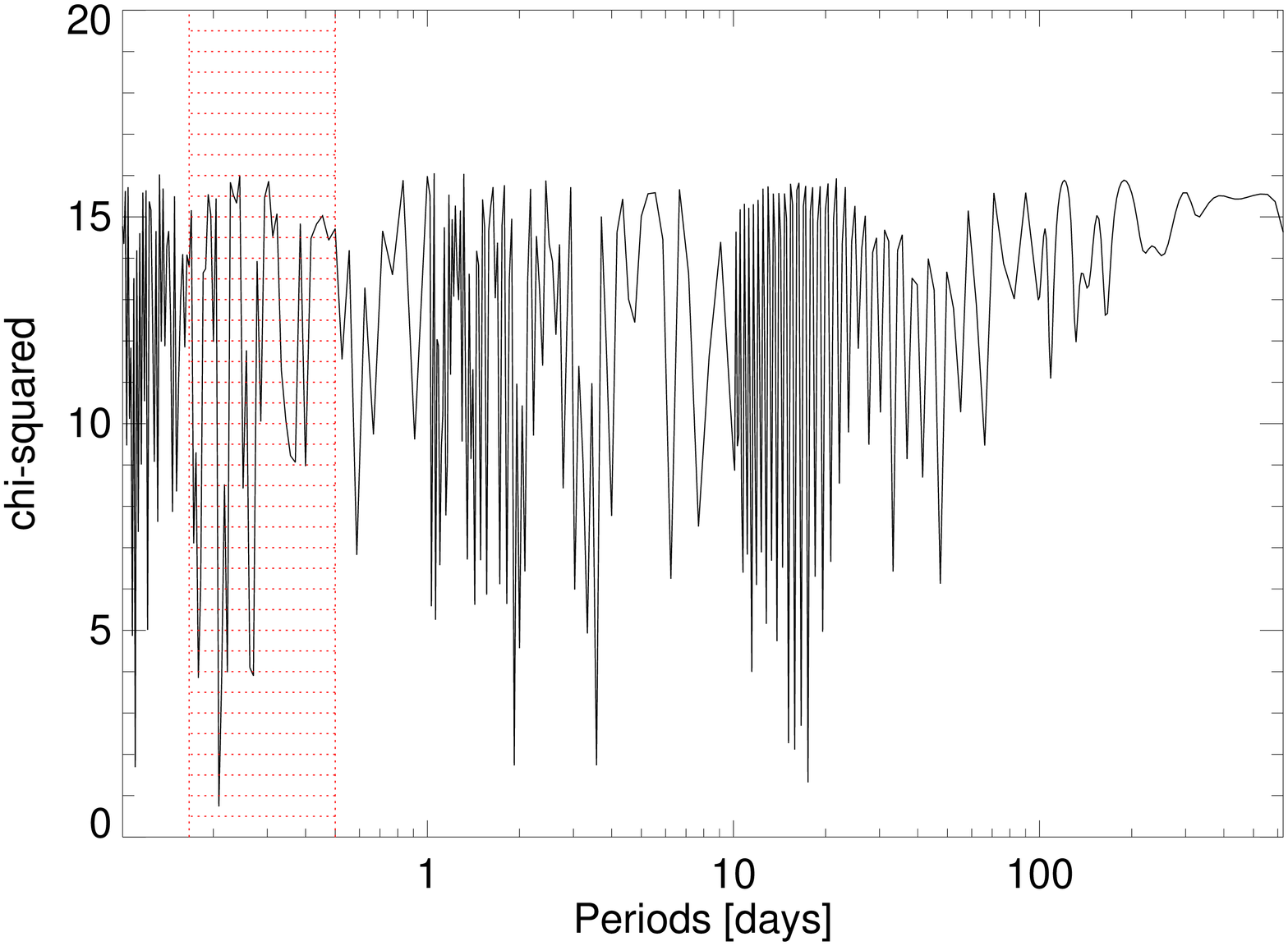}
\caption{Results of the $\chi^2$ sine fitting as function of trial frequencies. The red shaded areas show possible frequencies which might have been introduced by a moon. Top: TrES-1, bottom: TrES-2. \label{OCsin_chi2}}
\end{figure}

\begin{figure}
\centering
\includegraphics[angle=90,scale=0.33]{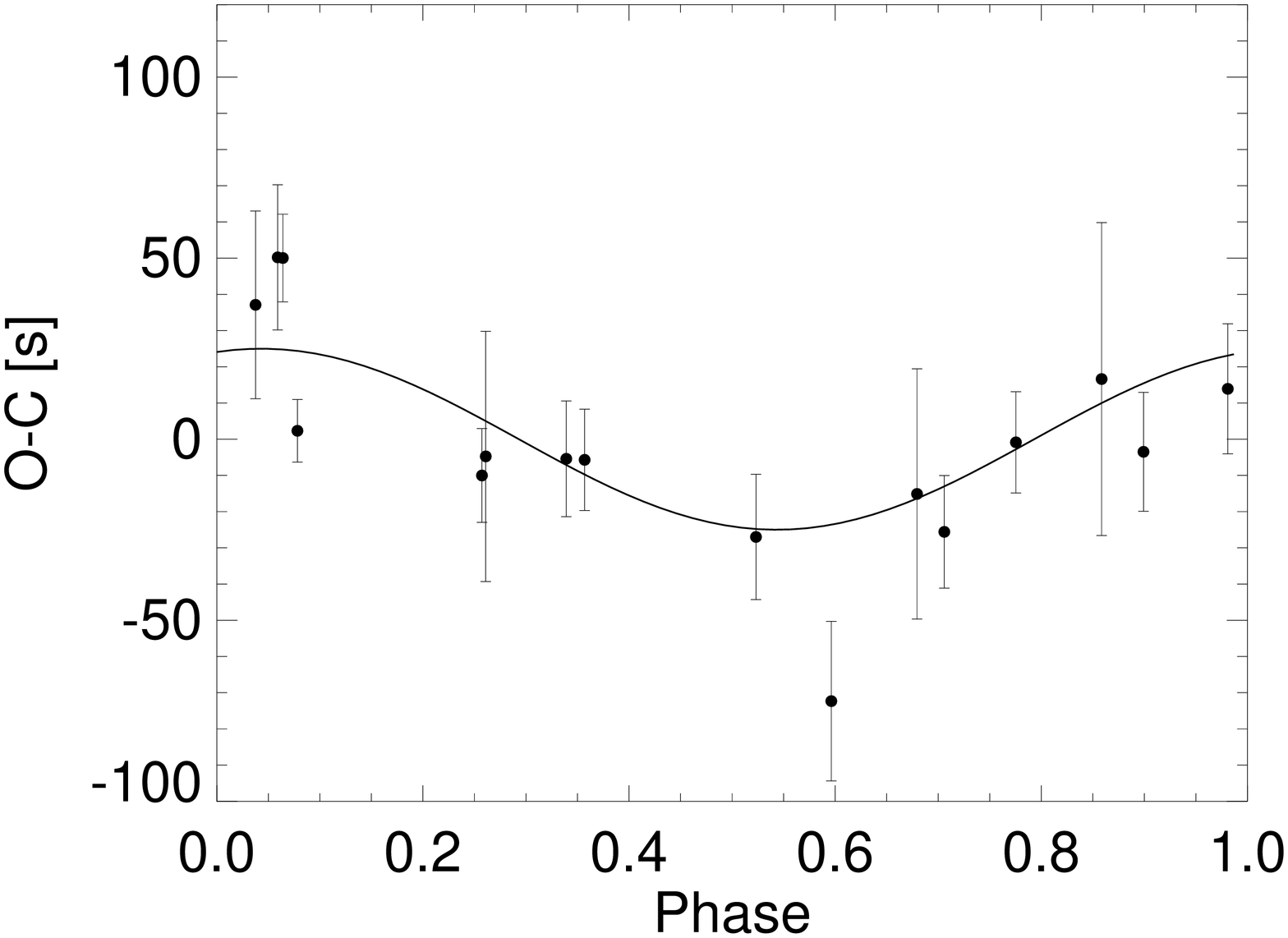}
\includegraphics[angle=90,scale=0.33]{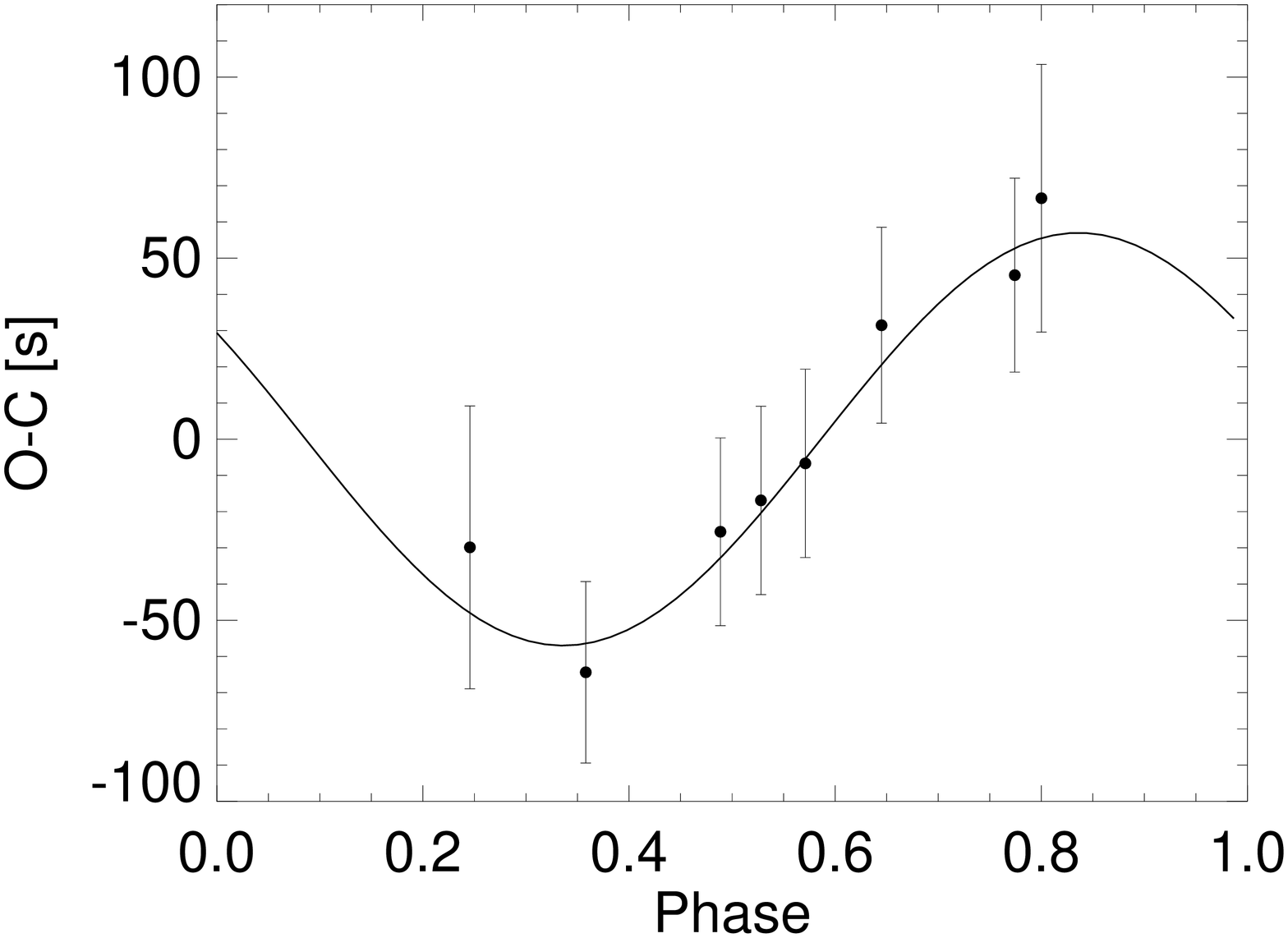}
\caption{Best fit sine function (solid line) and observed O-C values with error bars for TrES-1 and TrES-2, phased with period of that fit. This period is 0.06~cycles/day or 16.7 days for TrES-1 and 4.8~cycles/day or 0.21~days for TrES-2. \label{sinfit_best}}
\end{figure}

In order to search for any sinusoidal periodicity in the data with periods on the order of the campaign duration and shorter, we fitted amplitude and phase of a sinusoidal function to the O-C residuals after subtracting the linear fit with the parameters of Table \ref{bestfit_poly}. A sinusoidal transit timing variation might be caused by an exomoon around the transiting planet \citep{1999A&AS..134..553S,2004IAUS..213...80D,2007A&A...470..727S,2009MNRAS.392..181K}. For this fit we stepped through different trial periods between 1500 days (6.6 $\times$ 10$^{-4}$~cycl/d) and 0.1 days (10 cycl/d) for TrES-1 and for TrES-2 between 600 days (1.6 $\times$ 10$^{-3}$~cycl/d) and 0.1 days (10~cycl/d) in steps of 0.00001, 0.0001, 0.001, 0.01 and 0.1 cycl/d. For each trial period we then logged the $\chi^2$ residuals at the best fitting amplitude and phase.\\  
From Fig. \ref{OCsin_chi2}, showing the result of the $\chi^2$ sine fit, we see that we lack outstanding peaks of low $\chi^2$ values for TrES-1 and TrES-2, being it very unlikely that a real sinusoidal signal has more than one period. However, we also calculated the FAPs of the best sinusoidal fit against the linear polynomial, as described in Sect. \ref{intrinsic_period} and show the results in Table \ref{bestfit_poly} for comparison. We note that the best sinusoidal fit for TrES-1 with a period of 16.7 days and an amplitude of 25~s (Fig. \ref{sinfit_best}, upper graph) does not give a significant improvement against the linear polynomial, having a lower $\chi^2$ residual but a high FAP, but for TrES-2 we found a good sinusoidal fit with a FAP of 1 \% for a period of 0.2 days and an amplitude of 57~s (Fig. \ref{sinfit_best}, lower graph).
\begin{table*}
\caption[]{Comparison of different $\chi^2$ residuals for TrES-1 (upper values) and TrES-2 (lower values).}
\label{bestfit_poly}
\centering
\begin{tabular}{cccccccccc}
\hline\hline
\noalign{\smallskip}
 Fit         & $\chi^2$ &   $\kappa_0$  & $\kappa_1$  & $\kappa_2$  & $\kappa_3$  & $\nu$ & \multicolumn{3}{c}{False Alarm Probability vs.} \\
             &          &   s           & s           & $10^{-4}$~s & $10^{-6}$~s &       & original ephemeris & offset         & linear polynomial\\
\hline
TrES-1 \\
original ephemeris & 50.6 & 0.0 & 0.0      & -           &  -       & 15  & -    & -    & -    \\
offset          & 39.3 & -13.4 & 0.0      & -           &  -        & 14  & 0.065 & -    & -    \\
linear       & 28.0 & 9.5 & -0.1      & -           &  -            & 13  & 0.021 & 0.039 & -    \\
quadratic    & 23.7 & 6.3 &  0.1      & -3.6 $\pm$ 3.4  &  -        & 12  & 0.024 & 0.048 & 0.17 \\
cubic                  & 23.5 & 9.3 &  0.0      & 1.7      &  -0.7  & 11  & 0.058 & 0.12 & 0.38 \\
simulated best-fit O-C & 8.7     & -   & -         & -        & -      & 11    & -    & -    & 0.05     \\
simulated 1:2 MMR & 18.0 & -   & -         & -        & -      &  11   & -           & -    &  0.32 \\
simulated 2:3 MMR & 20.0 & -   & -         & -        & -      &  11   & -           & -    &  0.39   \\
simulated 3:2 MMR & 46.0 & -   & -         & -        & -      &  11   & -           & -    &  0.85 \\
simulated 5:3 MMR & 19.9 & -   & -         & -        & -      &  11   & -           & -    &  0.39 \\
simulated 2:1 MMR & 158.4 & -   & -         & -        & -      &  11   & -           & -    &  0.999 \\
best-fit sinusoidal fit &  21.5    & -   & -         & -        & -      &  12   & -           & -    & 0.38     \\
\hline

\noalign{\smallskip}

\hline
TrES-2\\
linear       & 13.1     & 28.9 &  -0.9 & -    & -           & 5 & - & - & -    \\
quadratic    & 12.6     & 24.8 &  -1.0 & 2.9 $\pm$ 21.7 & - & 4 & - & - & 0.72 \\
simulated best-fit O-C & 0.3 & -   & -         & -        & -      & 3    & -           & -    &   0.011  \\
simulated 1:2 MMR & 5.3     & -   & -         & -        & -      &   3  & -           & -    &  0.39    \\
simulated 2:3 MMR & 1.7     & -   & -         & -        & -      &   3  & -           & -    &   0.11   \\
simulated 5:3 MMR & 4.0     & -   & -         & -        & -      &   3  & -           & -    &   0.30   \\
simulated 2:1 MMR & 7.1     & -   & -         & -        & -      &   3  & -           & -    &   0.49   \\
simulated 3:1 MMR & 3.9     & -   & -         & -        & -      &   3  & -           & -    &   0.30   \\
best-fit sinusoidal fit &  0.7    & -   & -         & -        & -      &  4   & -           & -    & 0.011     \\
\hline
\end{tabular}
\end{table*}

%
%______________________________________________________________

\section{Discussion}
\label{concl}

For TrES-1 we obtained eight new O-C values and made use of 10 previously published values of which we removed two observations with a high probability of containing star spots. The standard deviation of the O-C values is $\sim$33~s and the maximum O-C deviation from the established ephemeris \citep{2007ApJ...657.1098W} is $\sim$60~s. Similarly, for TrES-2 we observed five transits and made use of three additional published mid-transit times \citep{2007ApJ...664.1185H}. The transit timing estimation at epoch 0 for TrES-2 had an error greater than 60 s and was hence removed. After correcting the ephemeris, the standard deviation of this O-C diagram is $\sim$43.9~s and the maximum O-C deviation, is 66.5 s.\\ 
In order to search for a light time effect, we fitted different polynomials. Assuming the validity of the quadratic function for TrES-1, we can use the quadratic term to estimate a possible acceleration and its corresponding 1-$\sigma$ error of $\frac{a_{\parallel}}{c}=-1.1 \pm 1.0 \times 10^{-14}$~s$^{-1}$, the negative value indicating a decelerating system. Comparing this deceleration to the acceleration for the solar system of a few $10^{-19}$~s$^{-1}$ \citep{2005AJ....130.1939Z} and CM Draconis of $\sim$10$^{-17}$~s$^{-1}$ \citep{2008A&A...480..563D}, the acceleration of TrES-1 could be several order of magnitudes larger, but is not well constrained given the current data. Following \citet{2008A&A...480..563D}, the minimum mass of a third body causing a given acceleration in dependence of its lateral distance $r_{\perp}$ is given by
\begin{equation}\label{mass}
\frac{m_{3}}{M_{\bigodot}} \geq 438.26 \left(\frac{a_{\parallel}}{\textrm{m s}^{-2}}\right)\left(\frac{r_{\perp}}{\textrm{AU}}\right)^2.\\
\end{equation}
We note that this is independent of the mass of the accelerating system. Using the distance to TrES-1 of 157 pc \citep{2004ApJ...616L.167S} and above value for $a_{\parallel}$, we obtain the minimum mass of a possible third body as a function of angular separation from TrES-1 of
\begin{equation}\label{mass_tres1}
\frac{m_{3}}{M_{\bigodot}} \geq 34 \left( \frac{\theta}{\textrm{arcsec}} \right)^2.\\
\end{equation}
This relation allows the identification of nearby objects found in any future high-resolution imaging; that is, any possibly found object has to fulfill Eq. \ref{mass_tres1} in order to be a potential source of the observed acceleration. For example, for a third object at a distance of 0.05$\arcsec$ the object's mass has to be at least 0.09~M$_\odot$ in order to cause the previously mentioned acceleration ($\frac{a_{\parallel}}{c}=-1.1 \pm 1.0 \times 10^{-14}$). The angular separation of 0.05$\arcsec$ translates into an orbit for a third object of 7.8~AU, corresponding to a period of 21.8~years. This stellar object would have caused a semi-amplitude in the radial velocity measurements of 1.0~km s$^{-1}$. However, the observation span of the available radial velocity measurements is only 49~days \citep{2004ApJ...613L.153A,2005ApJ...621.1072L}, which is 0.6~\% of the exemplary third object's period. Therefore, this object would not have been detected in existing radial velocity measurements. This example also shows that the obtained value for the acceleration is reasonable.\\
A second approach has been assuming perturbing objects at nearby orbits. Then for TrES-1, perturbers with a masses $>$1~M$_{\oplus}$ at the 2:3 and 2:1 MMRs are very unlikely due to the high $\chi^2$ residual with a high FAP against the linear fit. However, none of the simulated perturbers indicated a uniquely low $\chi^2$ peak. Therefore, we do not find support for any perturber, but we established upper mass limits above which we would obtain a FAP of 90 \% against the linear fit. For TrES-2, we obtained some low $\chi^2$ peaks; none of them in a low-order MMR. For the best-fit simulated O-C we obtained a perturber of 18~M$_\oplus$ and 0.049 AU. Due to the $\chi^2$ residuals below the one from the linear fit, none of the perturbers could be excluded and again we established upper mass limits for perturbers in the TrES-2 system. \\ 
\citet{2005MNRAS.364L..96S} did an analysis of transit times of TrES-1, but they found no convincing evidence for a second planet. We also found generally a higher $\chi^2$ near to the low-order MMRs. \citet{2005MNRAS.364L..96S} gave upper limits for additional planets in the system TrES-1, whereas we give additionally the best-fitting mass for any orbital distance. In general, their upper mass limit for additional planets decreases closer to the transiting planet, similar to our findings in Fig. \ref{bestmass}. We also found that additional planets with masses above 10~M$_{\oplus}$ at the MMRs are very unlikely, which is consistent with the results from \citet{2005MNRAS.364L..96S}, indicating that at the low-order MMRs and near zero eccentricity, the mass of the additional planet has to be below 10 M$_{\oplus}$.\\
Assuming sinusoidal transit timing variations, we note that we find a good O-C fit for TrES-2 with a FAP of 1.1 \%. However, since we find several good periods beyond the best one of 0.2 days, we conclude that we need more observations and maybe with higher precision in order to confirm any one of them as a possible exomoon. Whereas for TrES-1, we found no evidence that a sinusoidal function improved the fit significantly with respect to the linear one. \\ 
\citet{1999A&AS..134..553S}, \citet{2004IAUS..213...80D}, \citet{2007A&A...470..727S} and recently \citet{2009MNRAS.392..181K} discussed the possibility to detect moons around extrasolar planets using a timing offset induced by the wobble of the planet around the planet-moon barycenter. For a given timing offset $\delta t$ the mass of the possible moon $M_m$ can be estimated using \citep{1999A&AS..134..553S}:
\begin{equation}\label{moon_mass_e}
M_m \approx \frac{\pi a_P M_P}{a_M P_P} \delta t,\\
\end{equation}
where $a_P$ is the semi-major axis of the planet, $M_P$ the mass of the transiting planet, $a_M$ semi-major axis of the moon and $P_P$ the period of the planet. A moon around the transiting planet should have a semi-major axis which is between the Roche limit $r_{Roche}$ and the Hill radius $r_{Hill}$ \citep{2009MNRAS.392..181K}. The Roche limit is given by:
\begin{equation}\label{roche_limit}
r_{Roche}=R_P\left(2 \frac{\rho_P}{\rho_m} \right)^{1/3},
\end{equation}
where $\rho_p$ is the density of the planet and $\rho_m$ the density of the moon. As can be seen from Eq. \ref{roche_limit}, if $\rho_m > 2 \rho_p$, the Roche limit is inside the planet. This is most likely the case for a rocky moon around a gas giant planet. Therefore, we do not further consider the Roche limit as lower orbit limit, but the planetary radius. On the other hand, the Hill radius is given by:
\begin{equation}\label{hill_radius}
r_{Hill}=a_P\left(\frac{M_P}{3 M_*} \right)^{1/3},
\end{equation}
where $M_*$ is the mass of the star. Using the radius of the transiting planets, i. e. $R_{P,TrES-1}=1.081$~R$_J$ or $R_{P,TrES-2}=1.272$~R$_J$ and Eq. \ref{hill_radius}, we obtain possible orbital periods for a moon around the transiting planets of 4.2~h $<$ $P_m$ $<$ 15~h for TrES-1 and 3.9~h $<$ $P_m$ $<$ 12~h for TrES-2. If we consider only the corresponding period ranges in Fig. \ref{OCsin_chi2} (red shaded areas), we also find no clear peak there for TrES-1, but for TrES-2 the best peak of 0.2~days is in the range for an exomoon. Using Eq. \ref{moon_mass_e} we obtain a possible moon mass of 52~M$_\oplus$ for the best-fit amplitude of this peak.  \\
We note that \citet{2002ApJ...575.1087B} gave an analytical expression for an upper mass limit of possible moons. Generally, the more massive the moon the shorter its lifetime. \citet{1997ApJ...484..866L} estimated that close-in orbiting planets spin down into synchronous rotation very quickly. Therefore we use the procedure from \citet{2002ApJ...575.1087B} to estimate the upper mass limit of a moon around TrES-1 and TrES-2. We obtain an upper mass limit around TrES-1 on the order of $\sim$10$^{-6}$~M$_\oplus$ and around TrES-2 on the order of $\sim$10$^{-7}$~M$_\oplus$, similar to one obtained for HD 209\,458b by \citet{2002ApJ...575.1087B}. This means that moons with masses greater than $10^{-6}$~M$_\oplus$ will not have survived until now. This upper mass limit is clearly way below our detection limit for masses causing transit timing variations in the system TrES-1 and TrES-2. The closest moon at this upper mass limit would cause a timing amplitude of the order of 10$^{-6}$ s, which is not detectable even from space by several orders of magnitude. \\
Transit duration investigations on TrES-2 performed by \citet{2009arXiv0905.4030M} indicated a decrease of 3 min. between 2006 and 2008 in the duration. However, in our transit timing analysis, we do not find non-linear deviations of O-C times with a similar magnitude.
While we can not provide any firm detection, we can put upper mass limits which are consistent with our observations for TrES-1 and TrES-2, respectively. However, due to the gaped data, it might have been possible that we missed important points in the observed O-C diagram, like e. g. a transit timing measurement at the highest amplitude at the mean-motion resonances. We need at least about 3~years of continuous transit observations and with several transit observations per year in order to avoid the missing of windows of high-amplitude O-C deviations from perturbing bodies, with the maximum amplitudes being observable only during a few transit events, see Fig. \ref{OC_diag_tres62}. But even for low-mass perturbers such amplitudes should be easily measurable from ground at mean-motion resonances. On the other hand, outside the mean-motion resonances, we need higher precisions than obtainable from ground to detect the small amplitudes caused by a perturbing body of planetary mass, something that may be expected from the current satellite missions in operation. Any transit timing variation in TrES-2 may be expected to be confirmed in the near future by observations done with the Kepler Space Telescope.  \\

\begin{acknowledgements}
This article publishes observations made with the IAC-80 telescope operated by the Instituto de Astrof\'{i}sica de Canarias in the Observatorio del Teide. We thank the students and staff of the Teide Observatory and Angel de Vicente Garrido for his help with the IAC Supercomputing facility CONDOR. M. R. likes to thank the European Association for Research in Astronomy (EARA) for their support by a EARA - Marie Curie Early Stage Training fellowship. Part of this work was financed by grant ESP2007-65480-002-02 by the Spanish investigation and development ministry. We also thank Tristan Guillot and the anonymous referee for valuable comments that led to a significant improvement of the analysis presented here.
\end{acknowledgements}

\bibliographystyle{aa} % style aa.bst
\bibliography{bib_mar}

\end{document}